# A Formal Model of Distributed Systems For Test Generation Missions

Andrey A. Shchurov

*Department of Cybernetics, Faculty of Electrical Engineering,*
*Czech Technical University in Prague, The Czech Republic*

*Abstract*— Now a days, deployment of distributed systems sets high requirements for procedures and tools for the complex testing of these systems – virtualization and cloud technologies make another level of system complexity. As a possible solution, it is necessary to determine a formal list of control objectives – checklists. The automated generation of checklists involves analysing system models (with the analysis covering paths in a model). But complex distributed systems are usually a set of coexisting topologies which interact and depend on each other and it is necessary to use several models in order to cover different aspects. This work introduces a formal four layered model for test generation missions on the basis of the component-based approach and the concept of layered networks. The interlayer mapping determines how the topological properties on different layers affect each other and, as a consequence, represents technologies (virtualization, clustering, etc.) used to build distributed systems.

*Keywords*— Formal models, multi-layer graphs, distributed systems, automated test generation.

## I. INTRODUCTION

Computing systems have come a long way from a single processor to multiple distributed processors, from individual-separated systems to networked-integrated systems, and from small-scale programs to sharing of large-scale resources. As a consequence, nowadays the most difficult part of networked/distributed systems deployment is the question of assurance (whether the system will work) and verification. In the real world many systems have failed because their developers were under great financial and timing constraints and, as a consequence:

- had tested the wrong things;
- had tested the right things but in the wrong way;
- some things had been just simply forgotten and had not been tested.

As a possible solution, it is necessary to determine a formal list of control objectives during the design phase of the System Development Life Cycle (SDLC) and, as the next step, to show that each component of this list meets at least one test case during the implementation phase of the SDLC: i.e. it is necessary to have checklists.

The automated generation of checklists involves analysing system models (with the analysis covering paths in a model). In turn, these models face the great challenge of the nature of distributed systems:

- distributed systems tend to be complex – virtualization and cloud technologies make another level of system complexity;
- distributed systems tend to be heterogeneous – they usually include subsystems with very different characteristics.

On the other hand, graphs are powerful mathematical tools for modelling pairwise relationships among sets of objects/entities. But graphs traditionally capture only a single form of relationships between objects. However, complex heterogeneous systems rely on different forms of such relationships, which can be naturally represented by multi-layer graphs. Assuming that all layers are informative, they can provide complementary information. Thus, we can expect that a proper combination of the information contained in the different graph layers leads to the covering of the most important goals of distributed systems.

The rest of this paper is structured as follows. Section 2 introduces the related work. Section 3 presents the formal multilayer model of distributed systems for checklist generation missions. Section 4 introduces an example based on an industrial automation and control system. Finally, conclusion remarks and future research directions are given in Section 5.

## II. RELATED WORK

The most important goals of distributed systems are: (1) openness; (2) accessibility; (3) transparency; and (4) scalability [1]. Over the years a lot of effort has been invested in creating formal models that cover all these goals. However, each model typically represents only one aspect of the entire system. To evaluate the system as a whole, these models must be composed in such a way that their properties can be considered together. As a consequence, this composition has to:

- preserve the properties of each individual model;
- represent interaction between individual models.

Nowadays, hierarchical approaches for the modelling of complex distributed systems can be roughly classified into two categories:

- decomposition of complex models (tree structures);
- multi-layer (composed) models.

### A. Decomposition of Complex Models

Liu and Lee [2] and Eker et al [3] represent a structured approach – hierarchically heterogeneous. Using hierarchy, they can divide a complex model into a tree of nested submodels (see Fig. 1), which are at each level composed to form a network of interacting components (each of these networks to be locally homogeneous, while allowing different interaction mechanisms to be specified at different levels in the hierarchy). One key concept of hierarchical heterogeneity is that the interaction specified for a specific local network of components covers the flow of data as well as the flow of control between them.

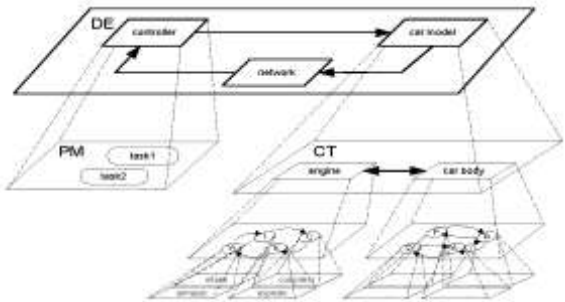

Fig. 1  A hierarchical model for the engine control systems [2].

The three dimensional analysis (Yadav et al [4]) decomposes a system structure into its physical elements and shows, in detail, how functional requirements can be fulfilled by individual product elements or groups of elements (see Fig. 2). The functional requirements propagate from the requirements for the complete product down to the elements in a hierarchical manner. The mapping between physical elements and functional requirements shows which physical elements have impact on the same function or which single element has an impact on different functions. The time dimension (or damage behaviour) helps in identifying which failure mechanisms have impact on physical elements and, as a consequence, on system functions.

Benz and Bohnert [5] define the Dependability Model as a model of use cases that are linked to system components they depend on. Dependability models have four levels: (1) user level; (2) function level; (3) service level; and (4) resource level. These models are constructed by identifying user cases or user interactions and then finding system functions, services and components which provide them. Once all system parts are found, the provision of use cases is modelled as links which show the dependability of user interactions on system components. Dependability models are shown either in a dependency table or in a dependency graph (see Fig. 3) to show the different dependencies between user interactions, system functions, services and system resources.

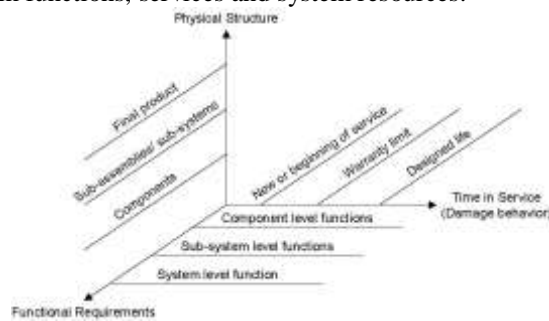

Fig. 2  Three dimensions of system structures [4].

### B. Page Layout

Kurant and Thiran [6] [7] introduce the multilayer model for studying complex systems. For simplicity, only a two-layer relationship is used (but the model can be extended to multi-layers). The lower-layer topology is called a physical graph and the upper-layer is called a logical graph (the physical and logical graphs can be directed or undirected, depending on the application). The number of nodes is equal for both layers. Every logical edge is mapped on the physical graph as a physical path. The set of paths corresponding to all logical edges is called mapping of the logical topology on the physical topology (see Fig. 4).

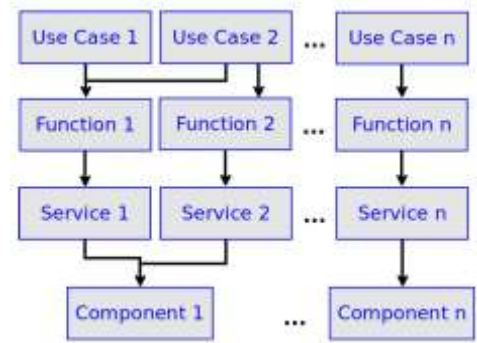

Fig. 3  Dependency graph [5].

Based on the multilayer model, Wong-Jiru [8] represents the Multi-Layer Model of Network Centric Operations (NCO). At each layer, the family of contributors is represented graphically as a network. The nodes represent individual contributors and the edges between them represent a layer-specific relationship. The model has five levels (see Fig. 5): (1) processes; (2) people; (3) applications; (4) systems; and (5) physical network. This layering scheme establishes a cohesive set of relationships for the major entities (people, processes and technologies) contributing to NCO. The layering hierarchy is based on the most direct interactions between major groups of entities.

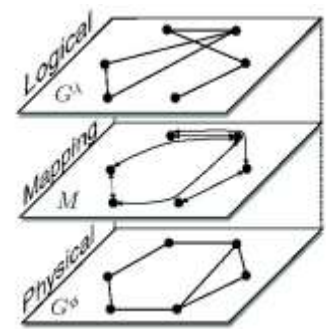

Fig. 4  Multilayer model [6] [7].

But in spite of the undoubted advantages of these models:
- network analysis metrics may be applied at any level, allowing each layer to be analyzed separately;

- mapping between layers allows the traceability of cause-and-effect;
- model components (layers, vertices, and edges) can be easily created at an appropriate level of abstraction (according to the rule "do not add more detail than is necessary");

they do not allow the representation of interlayer technologies (virtualization, clustering, replication, etc.) and the layered structures of real communication protocols (such as TCP/IP) are completely ignored.

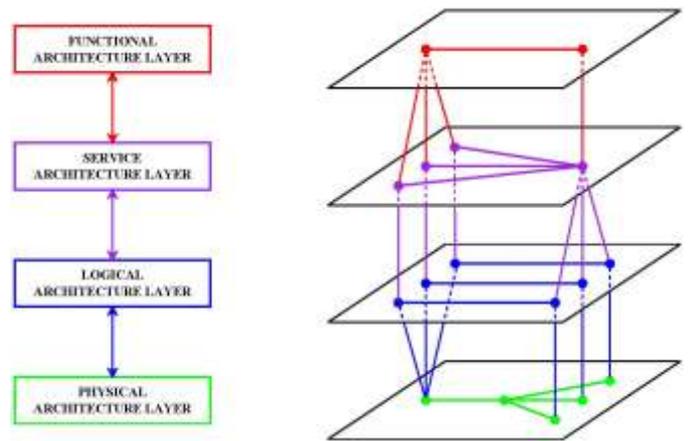

Fig. 6 Four layered model of distributed systems.

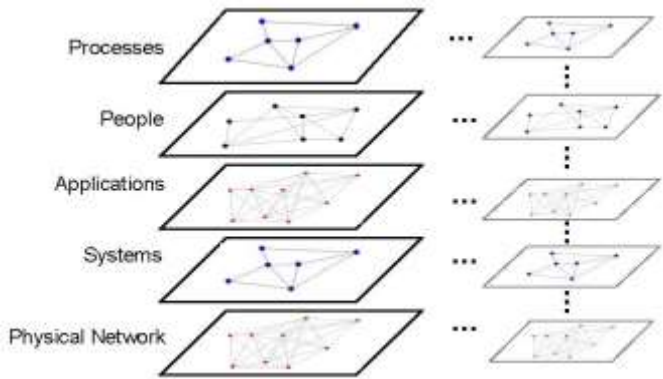

Fig. 5 Multi-layer model of Network Centric Operations [8].

III. FOUR LAYERED MODEL OF DISTRIBUTED SYSTEMS

The essential idea of our approach is based on two basic notions:

- component-based approach with its two important consequences: (1) components are built to be reused in different systems, and (2) component development process is separated from the system development process [2] [9];
- concept of layered complex networks [6].

The component-based approach refers to the fact that the functional usefulness of distributed systems does not depend on any particular part of these systems, but emerges from the way in which their components interact. Thus, the standard ISO/OSI Reference Model (OSI RM) [10] can be used as a starting point. But it cannot cover the all required aspects by itself (in practice, it is necessary to use several models in order to cover many different aspects). The necessary complements to OSI RM can be provided by the set of architectural models [11] as the most intuitive solution. In turn, the concept of layered complex networks ensures consistency between different models. And finally the graph theory (as a standard-de-facto) provides the necessary tools for models representations.

For our purposes the system model can be stated as a four-layered graph as follows (see Fig. 6):

- *The functional (or ready-for-use system) architecture layer* defines functional components and their interconnections − the enlarged viewpoint of end-users/customers. This layer is based on functional models [11]:
  - service-provider architectural model [11];
  - intranet/extranet architectural model [11];
  - single-tiered/multi-tiered architectural model [11];
  - end-to-end architectural model [11];

and covers the application (L7) layer of OSI RM.

- *The service architecture layer* defines software-based components (services/applications) and their interconnections. It is founded on flow-based models [11]:
  - client-server or centralized architectural model [1];
  - peer-to-peer or decentralized architectural model [1];
  - hybrid architectural model [1];

and covers the transport (L4), session (L5), presentation (L6) and partially the application (L7) layers – we cannot divide these layers in the case of commercial off-the-shelf (COTS) software.

- *The logical architecture layer* defines logical (virtual) components and their interconnections. It is based on topological models [11]:
  - LAN/MAN/WAN architectural model [12];
  - core/distribution/access architectural model [12];

and covers the network (L3) layer.

- *The physical architecture layer* defines hardware (physical) components and their interconnections. Like its predecessor, this layer is founded on topological models but covers the physical (L1) and data link (L2) layers – we cannot divide these layers in the case of COTS telecommunication/network equipment.

- *The interlayer projections* define all types of components hierarchical (interlayer) relations/mapping. These relations make the layered model consistent and represent interlayer technologies.

In this case, the model formal notation can be represented as:

$$G = (V, E, M)$$

where *G* is multi-layered 3D graph, derived from the system specification; *V(G)* is a finite, non-empty set of components (vertices); *E(G)* is a finite, non-empty set of component-to-component connections (horizontal edges); and *M(G)* is a finite, non-empty set of component-to-component interlayer mapping (vertical edges). Then, the system model $G_n$ for each layer *n* can be represented as a subgraph of *G*:

$$G_n = (V_n, E_n, M^n_{n-1}, V_{n-1})$$

And:

$$G = \bigcup_{n=1}^{N} G_n$$

where $V_n(G_n)$ is a finite, non-empty set of components on layer *n*; $E_n(G_n)$ is a finite, non-empty set of component-to-component connections on layer *n*; $M^n_{n-1}(G_n)$ is a finite set of component-to-component projection from layer *n* to layer *n-1*; and $V_{n-1}(G_n)$ is a finite set of components on layer *n-1*.

Generally, Gn is intransitive by default with the exception of the physical architecture layer. Each individual component of Vn (Gn) must have at least one top-down interlayer projection with the exception of the physical architecture layer. As a consequence:

$$|V_n(G_n)| \leq |M^n_{n-1}(G_n)|$$

In turn, $M^n_{n-1}(G_n)$ and $V_{n-1}(G_n)$ must be non-empty sets with the same exception – in this case, the definition of projection has no physical meaning.

In contrast to the multilayer model of complex systems [6], the sets of nodes of the system model on each layer are not identical:

$$V_n(G_n) \neq V_{n-1}(G_n)$$

The key factor is the arity of the component-to-component projection from layer n to layer n-1 (the top-down point of view). This parameter allows the technological solutions used to build the system to be represented:

- $N_n : 1_{n-1}$, e.g. virtualization technology representation (see Fig. 7 and Fig. 8);
- $1_n : N_{n-1}$, e.g. clustering technology representation (see Fig. 9);
- $1_n : 1_{n-1}$, e.g. a special case of dedicated components.

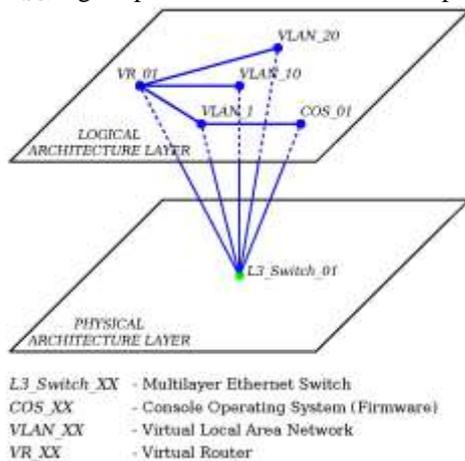

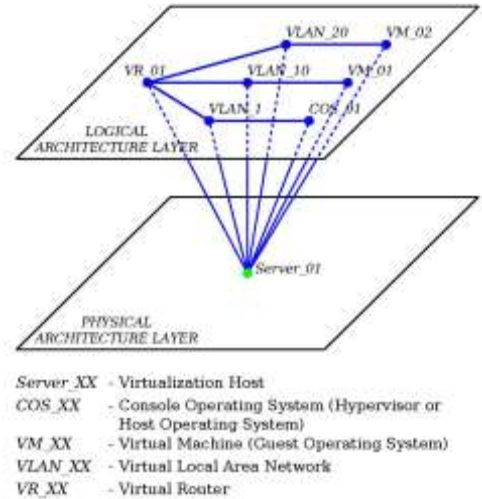

Fig. 7  Network virtualization example.

Fig. 8  Host virtualization example. The existence of virtual routers (VR) depends on implementation details: OpenStack [13] supports them, but VMware vSphere [14].

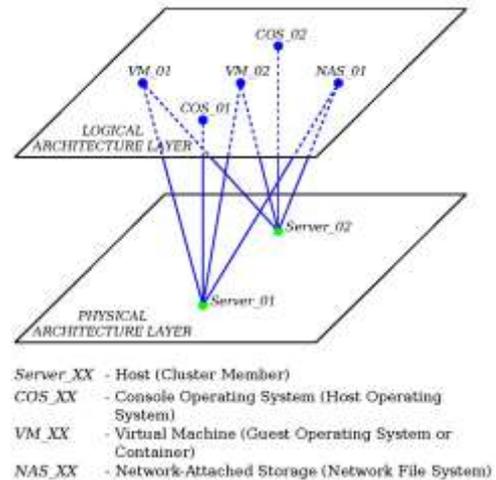

Fig. 9  Cluster example.

As the next step, it is necessary to determine the model requirements to cover the most important goals of distributed systems [1]:

- *Accessibility*. The main goal of a distributed system is to make it easy for the users (and services/applications) to access remote resources, and to share them in a controlled and efficient way. To support this function, the model must be consistent, i.e.:
  - Specifications (communication protocols and data representation formats) of components (nodes of the layered graph) must be compatible.
  - For each pair of individual components (defined by end-user requirement and/or specifications) which must communicate, a path (on the defined architectural layer) in the layered graph must exist.

- For each communication path (on the defined architectural layer) a top-down (lower layers) projection must exist (with the exception of the physical architecture layer).

- *Transparency.* An important goal of a distributed system is to hide the fact that its processes and resources are physically distributed across multiple computers. The concept of transparency can be applied to several aspects of a distributed system:
    - Access transparency. To support this function, the model must be consistent.
    - Location, migration and relocation transparency (the location of system resources). To support this function:
        - the model must be consistent;
        - the cardinality of the model must be equal to the maximal number of resource locations (nodes of the layered graph).
    - Concurrency and replication transparency. To support this function:
        - the model must be consistent;
        - the cardinality of the model must be equal to the maximal number of concurrent users and/or service replicas (nodes of the layered graph).
    - Failure transparency. To support this function, the model must remain consistent even if one (or more in special cases) arbitrary component is removed from this model, i.e. each communication path must have at least two top-down projections (a single point of failure does not exist).

- *Openness.* An open distributed system is a system that offers services according to standard rules that describe the syntax and semantics of those services. To support this function, specifications of system components (nodes of the layered graph) must be based on international standards.

- *Scalability.* Scalability of distributed systems is beyond the scope of this work – the process of adding more users and/or resources to the system usually requires the reconstruction of system models.

The experimental psychological work based on a large number of experiments related to sensory perception concludes that humans can process about 5–9 levels of complexity [15]. In this case we can state what the four-layered model might be simple enough for practical application.

## IV. CASE STUDY

As a practical example, we have an Industrial Automation and Control Systems (IACS) – the Converged Plantwide Ethernet (CPwE) Solution for Manufacturing Zone [16] with the exception of the server farm – in our case, the server farm is based on virtualization software VMware vSphere Standard Acceleration Kit with VMware Virtual Storage Appliance (VSA) as a shared storage [14].

In practical terms, we can define the object as a complex distributed system consisting of the following components:

- *Functional architecture layer.* From the viewpoint of end-users, the object can be described as a set of independent pairs provider/subscriber:
    - Industrial Automation and Control Systems as a Service (IACSaaS);
    - Network Management System as a Service (NMSaaS).

- *System architecture layer* (configuration details see [16] [17] [18]):
    - common network-based services – Active Directory (AD), Dynamic Host Configuration Protocol (DHCP), Dynamic Naming Services (DNS), Network Time Protocol (NTP);
    - network management services – CiscoWorks LAN Management Solution, Telnet/SSH;
    - virtualization infrastructure management services – vCenter and vSphere Update Manager;
    - manufacturing (automation and control) services – FactoryTalk Services and Applications;
    - client services – firmware, hypervisors, host and guest operating systems client services/applications (DNS-client, NTP-client, Web-browsers, etc.).

- *Logical architecture layer* (configuration details see [19] [20] [21]):
    - VLANs/subnets (include vSwitch configurations);
    - Virtual Routers (VRs) – independent and as part of the First Hop Redundancy Protocol (FHRP) configuration;
    - Virtual Machines (VMs) – Guest Operating Systems TCP/IP configurations;
    - Console Operating Systems (COSs) – firmware, hypervisors and host operating systems TCP/IP configurations.

- *Physical architecture layer.* Access, distribution, and core networking functions separate into distinct equipment (configuration details see [16] [19]):
    - L3/L2 (multilayer) and L2 Ethernet Switches;
    - Hosts – Servers and Workstations.

Table 1 shows the cardinality of the final 3D graph. This model was built as a set of PROLOG facts for test generation missions [22]. The requirements-coverage strategy application to the model covers:
- individual components;
- every interaction from the end-user requirements on functional, system, logical and physical architectural layers;

and, as a complement, checks the internal consistency of the system technical specifications with respect to the end-user requirements.

TABLE I
GRAPH (FOUR LAYERED MODEL) CARDINALITY

| Architectural layers | $G_n$ | | | | |
|---|---|---|---|---|---|
| | $n$ | $|V_n(G_n)|$ | $|E_n(G_n)|$ | $|M^n_{n-1}(G_n)|$ | $|V_{n-1}(G_n)|$ |
| Functional | 4 | 4 | 2 | 16 | 45 |
| System | 3 | 45 | 132 | 45 | 34 |
| Logical | 2 | 34 | 33 | 89 | 11 |
| Physical | 1 | 11 | 24 | - | - |

## V. CONCLUSIONS

Deployment of distributed systems sets high requirements for procedures, tools and approaches for complex testing of these systems. In this work we determined a formal model for test generation mission on the basis of the component based approach and the concept of layered networks. The model is a four layered 3D graph, derived from the system technical specifications, which covers all layers of OSI Reference Model and, as a consequence, both software-based and network-based aspects of distributed systems. In turn, the interlayer mapping (1) determines how the topological properties on different layers affect each other and, as a consequence, (2) represents technologies used to build the system (virtualization, clustering, replication, etc.).

Using this model and the graph theoretical metrics, both static and dynamic system analysis can be performed. The static analysis determined the characteristics of each layer based on the layer structure (or topology). It covers [22]:
- individual components;
- every interaction from the end-user requirements on functional, system, logical and physical architectural layers.

The dynamic analysis (or fault injection simulation) provides a means for understanding how distributed systems behave in the presence of faults. It includes [7]:
- successive removals of vertices/edges from the model (a layered 3D graph);
- impact assessments of those removals on the system internal consistency - disruption on an arbitrary layer might destroy a substantial part of the upper layers that are mapped on it, rendering the whole system useless in practice.

The strategy for the fault-injection experiments is generally based on methods for assessing the system reliability – the failure mode and effect analysis (FMEA). The typical FMEA is a document-centric evaluation, where a group of engineers evaluates the system [23]. But in the case of complex or non-standard systems, personal experience and/or intuition are often inadequate. As a consequence, future work will focus on the generation of FMEA reports for dependability testing missions using a dynamic analysis of formal layered models of complex distributed systems.

### ACKNOWLEDGMENT

This research has been performed within the scientific activities at the Department of Telecommunication Engineering of the Czech Technical University in Prague, Faculty of Electrical Engineering.